\documentclass[fleqn,usenatbib]{mnras}

\usepackage{newtxtext,newtxmath}
\usepackage[T1]{fontenc}
\usepackage[labelformat=empty]{subfig}
\DeclareRobustCommand{\VAN}[3]{#2}
\let\VANthebibliography\thebibliography
\def\thebibliography{\DeclareRobustCommand{\VAN}[3]{##3}\VANthebibliography}

\usepackage{graphicx}	% Including figure files
\usepackage{amsmath}	% Advanced maths commands

\usepackage{comment}	

\title[The sizes of massive satellites]{The size function of massive satellites from the $R_e-R_h$ and $M_{star}-M_h$ relations: constraining the role of environment.}

\author[The sizes of satellite MGs]{
L. Zanisi,$^{1}$\thanks{E-mail: l.zanisi@soton.ac.uk}
F. Shankar,$^{1}$, M. Bernardi$^{2}$, S. Mei$^{3}$, M. Huertas-Company$^{4}$
\\
$^{1}$Department of Physics and Astronomy, University of Southampton, B46 University Road, SO17 1BJ, Southampton, UK \\
$^{2}$  Department of Physics and Astronomy, University of Pennsylvania, Philadelphia, PA 19104, USA\\
$^{3}$Universit\'e de Paris, CNRS, Astroparticule et Cosmologie, F-75013 Paris, France\\
$^{4}$Instituto de Astrof\'isica de Canarias (IAC); Departamento de Astrof\'isica, Universidad de La Laguna (ULL), E-38200, La Laguna, Spain
}

\date{Accepted for publication in MNRAS}

\begin{document}
\label{firstpage}
\pagerange{\pageref{firstpage}--\pageref{lastpage}}
\maketitle

% Abstract of the paper
\begin{abstract}
In previous work we showed that a semi-empirical model in which galaxies in host dark matter haloes are assigned stellar masses via a stellar mass-halo mass (SMHM) relation and sizes ($R_e$) via a linear and tight $R_e-R_h$ relation, can faithfully reproduce the size function of local SDSS central galaxies and the strong size evolution of massive galaxies (MGs, $M_{\rm star}>10^{11.2}M_\odot$). In this third paper of the series, we focus on the population of satellite MGs. We find that without any additional calibration and irrespective of the exact SMHM relation, fraction of quenched galaxies or level of stellar stripping, the same model is able to reproduce the local size function of quiescent satellite MGs in SDSS. In addition, the same model can reproduce the puzzling weak dependence of mean size on host halo mass for both central and satellite galaxies. The model also matches the size function of starforming satellite MGs, after assuming that some of them transform into massive lenticulars in a few Gyr after infalling in the group/cluster environment. However, the vast majority of satellite lenticulars is predicted to form before infall. The $R_e-R_h$ appears to be fundamental to connect galaxies and their host haloes. 
\end{abstract}

% Select between one and six entries from the list of approved keywords.
% Don't make up new ones.

%%%%%%%%%%%%%%%%%%%%%%%%%%%%%%%%%%%%%%%%%%%%%%%%%%

%%%%%%%%%%%%%%%%% BODY OF PAPER %%%%%%%%%%%%%%%%%%

\section{Introduction}
\label{sec:introduction}
In the $\Lambda$CDM cosmogony, galaxies form and evolve in dark matter haloes \citep{White&Rees1978}. Thus, it is believed that the scaling relations between galaxy and host halo properties are a crucial probe of galaxy evolution. In particular, the relationship between galaxy stellar mass and host dark matter halo mass (the stellar-mass-halo-mass relation, SMHM) has attracted much attention in the past decade \citep{Leauthaud+12,Behroozi+13,Shankar+14_SMHM, Rodriguez-Puebla+17,Lapi+18_ETGs}. In addition, several works suggest empirical evidence for a tight and universal galaxy effective radius-halo virial radius (i.e., $R_e-R_h$) relation \citep{Kravtsov2013, Huang+17, Somerville+18, Lapi+18_disks}, which holds for both starforming and quenched galaxies, with a different normalization but similar scatter \citep{Zanisi+20}.
 
\citet{Stringer+14} outlined a semi-empirical model where dark matter haloes obtained from N-body simulations are populated with galaxies of a given size and stellar mass, using the SMHM and the $R_e-R_h$ relations. We applied this framework to central galaxies in two recent papers (\citealt{Zanisi+20,Zanisi+21_MGs}, Paper I and Paper II respectively). In Paper I we characterized the normalization and dispersion of the $R_e-R_h$ relation for nearby (i.e., $z\sim 0.1$) central galaxies. The inferred scatter of the $R_e-R_h$ relation was found to be as small as $0.1$ dex for Massive Galaxies (MGs, $M_{\rm star}>10^{11.2}M_\odot$), which suggests a strong link between galaxy size and halo virial radius in this stellar mass regime. In Paper II we found that a constant $R_e-R_h$ relation is able to explain the puzzling steady size increase of MGs up to $z\sim 3$. The relative proportion of compact MGs (irrespective of the definition of compactness, see e.g. \citealt{Damjanov+15_numbers_compacts})  was found to be a strong function of the shape and scatter of the SMHM. %As shown in Paper 1 and further emphasized in Paper 2, the latter two findings are a consequence of two simple facts: 
% \begin{itemize}
%     \item The size of dark matter haloes of a given mass progressively increases as redshift decreases
%     \item The SMHM maps MGs in progressively more massive haloes, with the conditional halo mass function (CHMF, $\phi(M_h|M_{\rm star})$), exhibiting a strong dependence on the specific SMHM relation adopted. This dependence is reflected in the size function by means of the $R_e-R_h$ relation.  
% \end{itemize}

 While in Papers I and II we calibrated and deployed the semi-empirical framework described above only for central galaxies, in this paper we will focus on the population of \emph{satellite} MGs. In particular, we will mainly focus on the following two still open issues: 
 \begin{itemize}
 \vspace{-0.5em}
     \item The environmental dependence of the sizes of MGs. Previous work (e.g., \citealt{Huertas-Company+13_environment,Shankar+14_sizes,Sonnefeld+19}) showed that, at the present cosmic time, the sizes of all MGs should not differ, in terms of a  ``mass-normalised'' size  ({$\log \gamma=\log R_e+0.83(11-\log M_{star}$), which cancels the dependence on underlying differences in stellar mass on the average size difference}, e.g., \citealt{Huertas-Company+13_environment}), by more than $\Delta \gamma \lesssim 30-40$\% between low-mass groups and massive clusters. We will test this trend within the framework of the $R_e-R_h$ relation.
     %but the referee might ask, why do you do it only for satellites and not for centrals?!?
     %within  as a testbed of our semi-empirical framework. 
     \item The formation of lenticular (S0) galaxies within the cluster environment (e.g., \citealt{Bekki&Couch2011}). %The morphology and star formation activity of star forming galaxies are believed to be profoundly affected in rich environments. Several concurring, yet so far poorly constrained, phoenomena that occur in groups and clusters (e.g., \citealt{Boselli&Gavazzi2006,Peng+10,Bahe+17_hydrangea,DeLucia+19_satellites}) are believed to suppress the star formation activity of star forming galaxies as well as induce a morphological transformation towards the early type sequence (e.g., \citealt{Smith+05_morphodensitysincez1,Deeley+20_S0formation_SAMI,Coccato+20_S0s}). 
     We aim to probe the viability of a model in which massive spirals transform into massive S0s over a given timescale (e.g., \citealt{Smith+05_morphodensitysincez1,Deeley+20_S0formation_SAMI}). 
\end{itemize} 
 
 This Letter is organised as follows. We describe the comparison data in Section \ref{sec:data}, the modelling in Section \ref{sec:methods} and the main results in Section \ref{sec:results}. We give final remarks in Section \ref{sec:discussion}.

\section{Data}
\label{sec:data}
Our reference data is the Sloan Digital Sky Survey (SDSS) Data Release 7 \citep{Abazajian+09} as presented in \citet{Meert+15,Meert+16}. We use the best-fitting  \texttt{S\'ersic-Exponential} or \texttt{S\'ersic} profile to r-band observations, and adopt the mass-to-light ratios from \citet{Mendel+14}.% Truncation of the light profile becomes an important factor in the computation of galaxy sizes $R_e$. Without truncation, the sizes of galaxies which are fit by one component are overestimated, which results in highly skewed galaxy size distributions. Therefore, in this paper 
We adopt the truncation of the light profile as prescribed in \citet{Fischer+17_truncation}. We also match the Meert et al. catalogs with both the \citet{Yang+12_groups} group catalogs, {complete in the halo mass range of interest here}, which allows us to identify central and satellite galaxies, and the \citet{Dominguez-Sanchez+18} deep-learning based morphological catalog. {The matching, we checked, preserves full completeness in the stellar and halo mass functions of MGs.} We use TType<=0 to select Early Type Galaxies, which we assume to be quenched (Massive Quenched Galaxies, MQGs), and TType>0 for late type galaxies, which we assume to be star forming (Massive Star Forming Galaxies, MSFGs). We also consider massive S0 galaxies (MS0s), which in the \citet{Dominguez-Sanchez+18} catalog are identified with a probability of being S0, $P_{S0}$. Using standard $V_{max}$ weighting, we produce the size function $\phi(R_e)$ of central and satellite MQGs and MSFGs, which are the targets for our model to reproduce. The size function of MS0s is produced by weighting the MQGs size function by $P_{S0}$. The errorbars are computed using jackknife resampling.

\section{Methods}
\label{sec:methods}

We here closely follow the modelling approach of %\citet{Zanisi+20}, Zanisi et al. submitted
Papers I and II\footnote{We make extensive use of the \texttt{COLOSSUS} Python package \citep{DiemerColossus}.}: 
\begin{enumerate}
\item \textbf{Dark matter catalogues.} %Dark matter haloes constitute the starting point of our model.
We start from the publicly available\footnote{\url{https://www.cosmosim.org/cms/simulations/mdpl/}} data products from the MultiDark-Planck (MDPL) simulation  \citep{Klypin+16_Multidark}. 
%from the MultiDark project \citep{Prada+12,Riebe+13_MultiDark_datarelease}. The MDPL simulation consists of 3048$^3$ dark matter particles evolved  with the \texttt{L-GADGET-2} code (based on \textbf{springel2005} in a cosmological box of 1 Gpc/h a size and with the \textbf{planck2013} cosmological parameters. 
For (unmerged) subhaloes at $z\sim 0.1$ 
%, which may undergo significant stripping \citep{Wu+12_subhalostripping_Vpeak} from the cluster potential, 
we adopt the peak virial mass $M_{peak}$ attained during their mass assembly history, before accretion. %{\color{green}For the remainder of this paper, we will only consider satellites that have survived down to the redshift of observation, $z\sim 0.1$, i.e. mergers between satellite and central galaxies are not modeled.}
Dark matter halo masses $M_h$ are defined as virial overdensities within a radius $R_h$ \citep{Bryan&Norman98}.
%\begin{equation}
%R_{h}=\Bigl( \frac{3M_{h}}{4\pi \cdot \Delta\rho_{\Delta}} \Bigr)^{\frac{1}{3}}
%\label{eq:Rhalo}
%\end{equation}
% where $\Delta$ is the virial overdensity with respect to the cosmological critical density $\rho_{\Delta}$. 
 
\item \textbf{The SMHM relation.} We model the link between galaxies and dark matter via the SMHM, which is a monotonically increasing function of halo mass \citep[e.g.,][]{Vale&Ostriker2006,Shankar+06}. We further assume the SMHM to follow a Gaussian distribution in $\log{M_{\rm star}}$ at fixed halo mass, $\sigma_{\rm SMHM}=0.15$dex \citep[][]{Grylls+20_satellites}. 
%\begin{multline}
% SMHM(M_{\rm star}|M_{\rm h},z) =\\ 
% \frac{1}{\sqrt{2\pi \sigma_{SMHM}^{2}}} exp \Bigl [ -  \frac{(logM_{\rm star}(M_{\rm h},z)-\langle logM_{\rm star}(M_{\rm h},z) \rangle)^2} {2\sigma_{SMHM}^2} \Bigr].
%\label{eq:SMHM}
%\end{multline}
As our benchmark SMHM relation, we assume the $z\sim 0.1$ SMHM relation by \citet{Grylls+20_satellites}, which was obtained by fitting the \citet{Bernardi+17_highmassend} 'PyMorph' \texttt{SerExp} stellar mass function (SMF). The stellar mass of \citet{Bernardi+17_highmassend} are obtained without the truncation of the light profile (e.g., \citealt{Fischer+17_truncation}). However, the truncation adopted in this work (see Section \ref{sec:data}) results in the high mass end of the SMF being slightly less populated, requiring fine-tuning in two of the \citet{Grylls+20_satellites} parameters, namely $\gamma_0\approx0.57$ and $M_{10}\approx11.95$.

\item \textbf{The $R_e-R_h$ relation.} We assign a half light radius $R_e$ to each galaxy according to the ansatz:
\begin{equation}
R_e = A_k R_h,
\label{eq:K13model}
\end{equation}
which is based on the empirical findings by \citet{Kravtsov2013}, and that we call the \emph{K13 model}. 
Here $A_k$ is the normalization which in principle may vary galaxy stellar mass and/or morphology (e.g., \citealt{Huang+17}). 
\item \textbf{Star formation activity.} MGs are a bimodal population in terms of star formation activity. 
 \begin{itemize}
     \item  \textbf{Relative fraction of MSFGs and MQGs}.  In Paper II we modelled explicitly the probability of a central galaxy being quiescent by adopting the parametrization by \citet{Rodriguez-Puebla+15}, which is valid at $z\sim0.1$: 
    \begin{equation}
    f_{Quench} (M_h)= \frac{1}{b_0 + [\mathcal{M}_0 \times10^{12}/M_h (M_{\odot})]},
    \label{eq:fQevol}
    \end{equation}
    where $b_0=1$. $f_{Quench}$ is a monotonically increasing function of halo mass, with a characteristic mass scale $\mathcal{M}_0$ above (below) which more (less) than 50\% of galaxies are quiescent (star forming). We further assumed that $\mathcal{M}_0$ evolves as $\mathcal{M} (z) = \mathcal{M}_0 + (1+z)^\mu$
    %\begin{equation}
    %\label{eq:betaz}
    %\mathcal{M} (z) = \mathcal{M}_0 + (1+z)^\mu,
    %\end{equation}
    which qualitatively reproduces the quenching downsizing phenomenon (e.g., \citealt{Behroozi+18}).   Here $\mu>0$ is a free parameter that controls the quenched fraction in dark matter haloes at a given cosmic time. By using current estimates of the galaxy stellar mass function for quenched and star forming galaxies (\citealt{Davidzon+17}, \citealt{McLeod+20}), in Paper 2, we have found that $\mu$ is likely to lie at intermediate values, $\mu\approx 2.5$. We will use $\mu=2.5$ as a reference but we will explore other possible values in the next Section. Following paper 2, we also set $\mathcal{M}_0$=1.5. It is important to note that \emph{the $f_{Quench}$ model applies strictly only to central galaxies}. In particular, satellites have their star formation activity set at a time when they were central, $z_{peak}$, but may be modified by environmental effects (see bullet \texttt{v}).
    \item \textbf{The sizes of MSFGs and MQGs} The effective radii of MSFGs are observed to be higher than those of MQGs by a roughly constant factor \citep{Mowla+18}. We capture this feature by assuming that the normalization of the K13 model differs for MSFGs and MQGs. Likewise, we will assume that the scatter of the K13 model differs for MSFGs ($\sigma_{K,SF}$)and MQGs ($\sigma_{K,Q}$). Following Paper II, we calibrate $A_{K,SF}, \sigma_{K,SF}$ and $A_{K,Q}, \sigma_{K,Q}$ on the size function of central galaxies at $z\sim 0.1$ (see Figure~\ref{fig:sizefunction_frozen}).
    %{\color{red}following Appendix A in \textbf{Zanisi+21 submitted}} (see Figure \ref{fig:sizefunction_frozen}). %{\color{red} We find ($A_{K,SF}, \sigma_{K,SF}$, $A_{K,Q}, \sigma_{K,Q}$) = (0.024,0.13 dex, 0.014,0.10 dex).}
    %we don't need to repeat this info that is contained in the caption of Figure 1, it saves space.
    \end{itemize}
\item \textbf{Environmental effects.} In what follows we will include models with stellar stripping 
following the results from the N-body simulations by \citet{Smith+16_stripping}, who 
%found that the stripping of stars is significant only after a fraction $f_{\rm DM} \sim80$\% of the subhalo dark matter mass is stripped. The resulting 
suggest a mass loss given by $M_{\rm star,strip}/M_{\rm star}= \exp[1-14.2f_{\rm DM}]$, with $f_{\rm DM}$ the dark matter fraction. We then update the sizes following \citet{Shankar+14_sizes} and \citet{Hearin+19}, who assume that $R_e$ decreases proportionally to the decrease in stellar mass along the $R_e-M_{\rm star}$ relation.
We will also consider models in which some MSFGs are quenched and morphologically transformed in S0 galaxies by the environment (e.g., \citealt{Smith+05_morphodensitysincez1}) over a given timescale $\Delta T_{\rm transf}$.
\end{enumerate}

%In summary, we initialise dark matter subhaloes at the peak of their mass accretion history (or at $z\sim0.1$ for distinct haloes) with galaxies whose stellar mass, size and star formation activity are set by the SMHM relation, the K13 relation and the $f_{Quench}$ model respectively. Satellites may undergo further environmental effects after infall. The data for central galaxies are used to calibrate the free parameters of the model, while the model outcome for satellite galaxies constitutes a prediction. 

%\section{Frozen \& stationary model}
\section{The local size function of starforming and quenched massive satellite galaxies}
\label{sec:results}

\begin{figure*}
    \subfloat[]{{\includegraphics[width=0.8\textwidth]{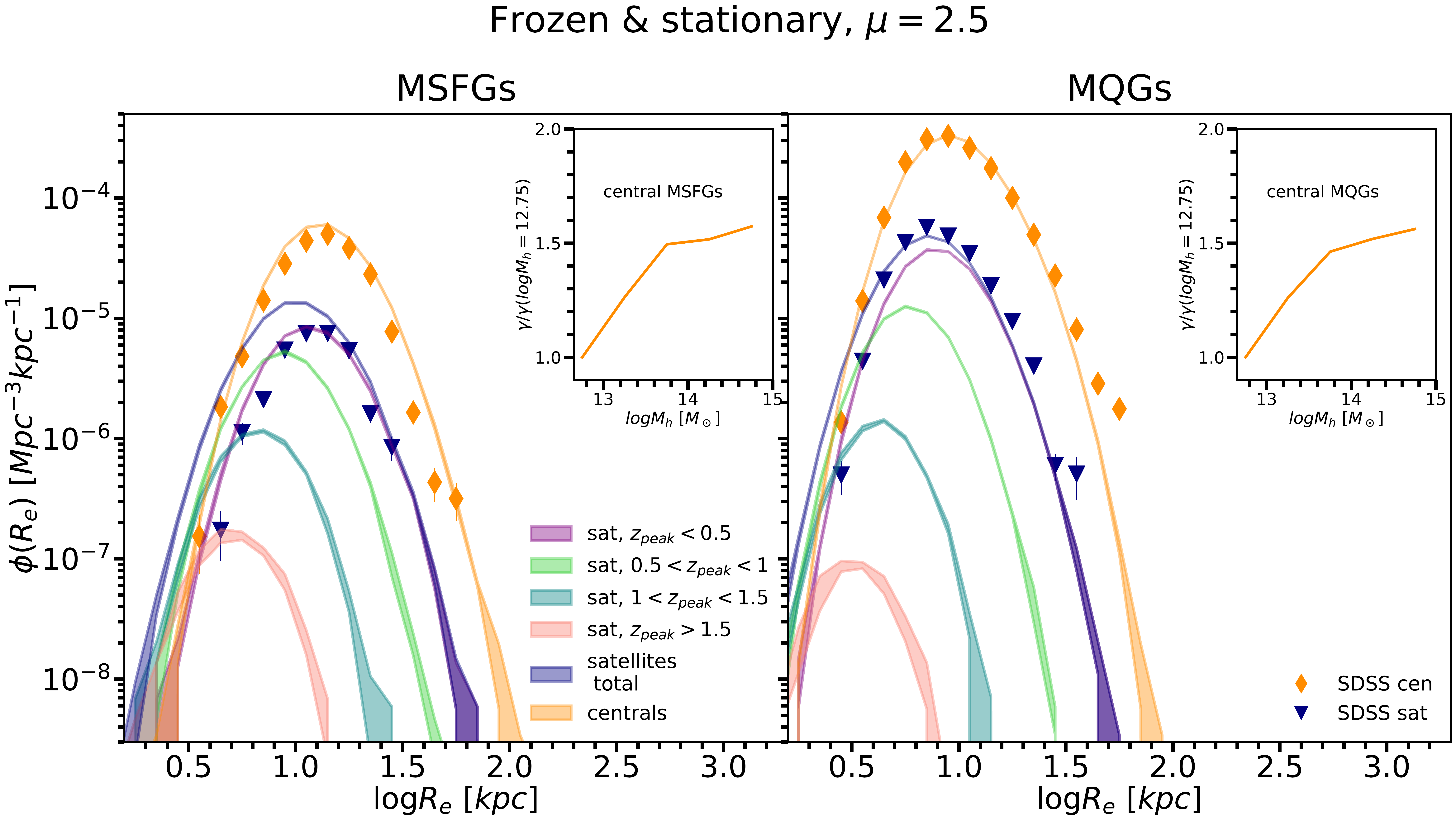}}}
    \caption{Size function of MSFGs (left) and MQGs (right) for SDSS central (orange diamonds) and satellites (blue triangles). The solid lines and filled regions show our ``frozen \& stationary" model, i.e. satellites do not evolve after infall and the SMHM relation (adapted from \citealt{Grylls+20_satellites}, see Section \ref{sec:methods}) is taken at $z\sim0.1$ and it is assumed not to evolve at high redshift. We calibrate the free parameters of the model on the size functions of central galaxies: $\sigma_{K,SF}$=0.13 dex, $A_{K,SF}$=0.024, $\sigma_{K,Q}$=0.10 dex, $A_{K,Q}$=0.013 and $\mathcal{M}_0$=1.5. The blue filled regions indicate the total surviving population of satellites accreted by $z\sim 0.1$.} %Although the model works well at first order, it is clear that the low-size end of satellite MSFGs size function is overestimated, and the sizes of satellite MQGs are systematically underestimated by roughly $0.05$ dex. The magenta, green, aquamarine and pink regions show the size functions of satellites initialised at $z_{peak}<0.5$, $0.5<z_{peak}<1$, $1<z_{peak}<1.5$ and $z_{peak}>1.5$, and their sum corresponds to the blue filled regions. These lines are a function of the precise implementation of redshift evolution of the model ingredients.}
    \label{fig:sizefunction_frozen}
\end{figure*}

\begin{figure*}
    \centering
    \includegraphics[width=0.95\textwidth]{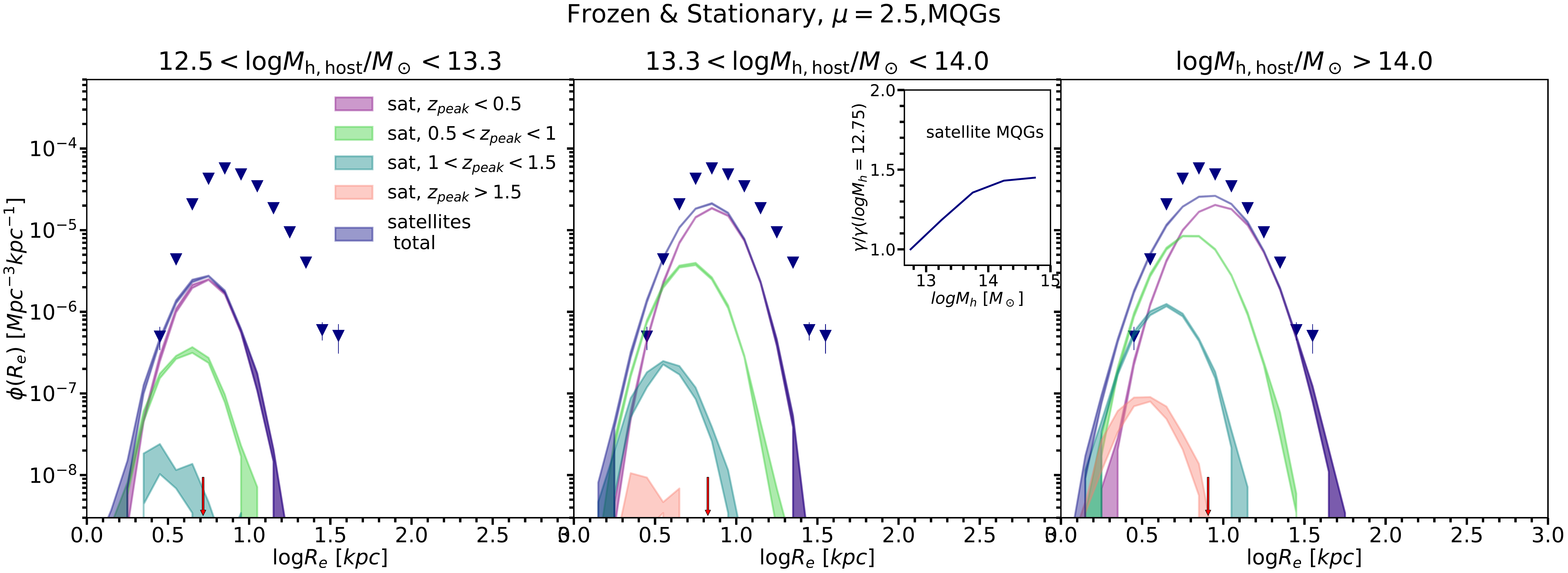}
    \caption{The size functions of satellite MQGs from the "frozen \& stationary" model, computed in different bins of halo mass corresponding to low-mass groups (left panel), groups and low-mass clusters (central panel) and massive clusters (right panel). The mean sizes in each environment are shown with an arrow. Lines are as in Figure \ref{fig:sizefunction_frozen} but here they refer to each halo mass bin, and symbols refer to the total SDSS satellite MQGs size function.}
    \label{fig:sizefunction_frozen_halobins}
\end{figure*}

%\begin{figure*}
%    \centering
%    \includegraphics[width=0.95\textwidth]{Sizefunct_HaloMassRange_11.6_MPeak_gamma10_0.57_gamma11_0_beta11_0_M10_11.95_M11_0_SHMnorm11_0_MSFGs_mu2.5_benchmark.pdf}
%    \caption{\textbf{Not sure we should keep this figure unless we want to discuss environmental dependence of S0s}. The size functions of satellite MSFGs from the "frozen \& stationary" model, computed in different bins of halo mass corresponding to low-mass groups (left panel), groups and low-mass clusters (central panel) and massive clusters (right panel). Lines are as in Figure \ref{fig:sizefunction_frozen} but here they refer to each halo mass bin, and symbols refer to the total SDSS satellite MQGs size function. }
%    \label{fig:sizefunction_frozen_halobins_MSFGs}
%\end{figure*}

\begin{figure*}
    \centering
    \includegraphics[width=0.8 \textwidth]{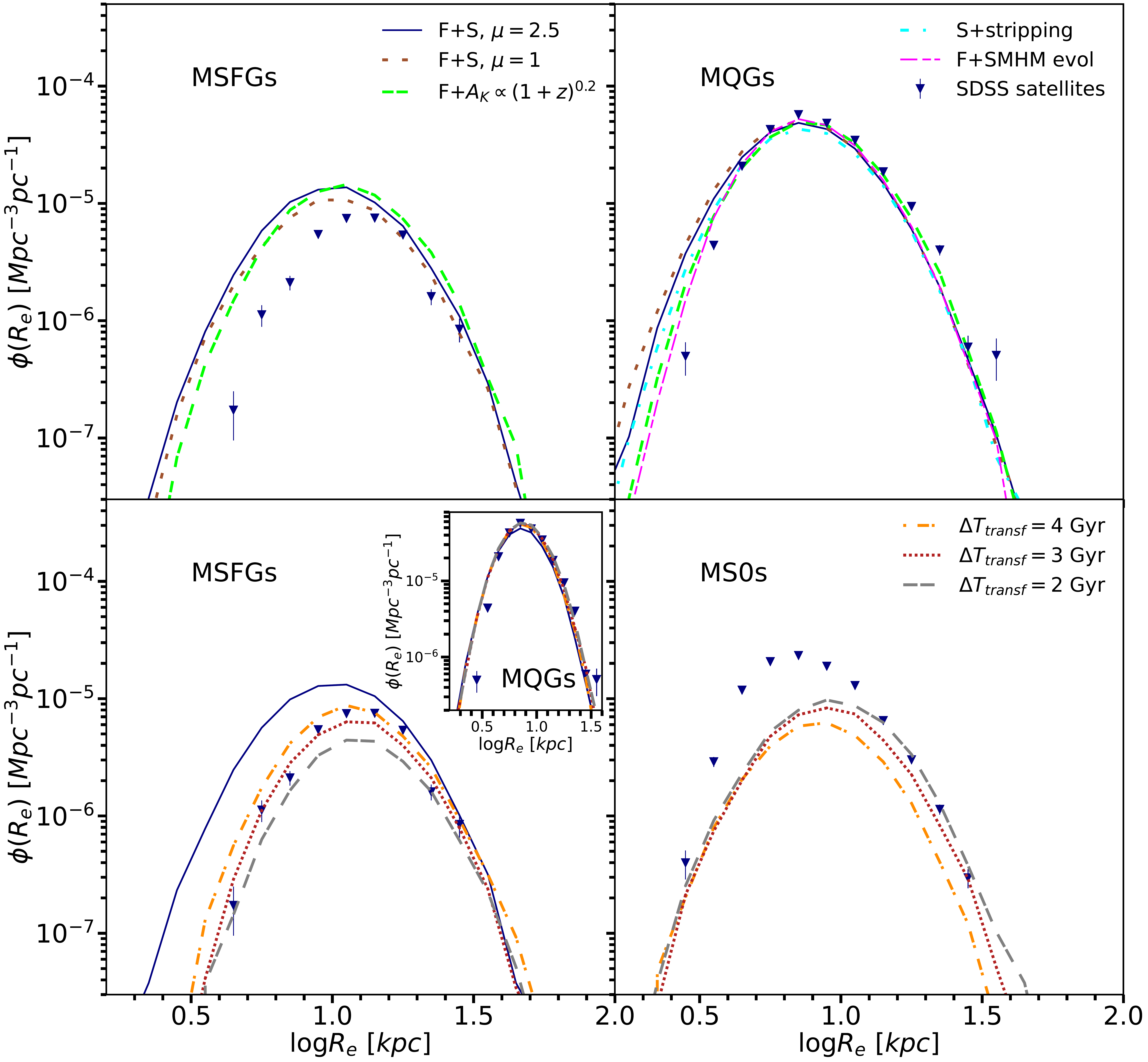}
    \caption{\emph{Top row}: Variants to the benchmark "Frozen \& Stationary" (F+S) model (solid blue lines). The dotted brown lines show a model with $\mu=1$. In the right panel, the cyan dot-dashed lines show the result of a model with stellar stripping. The long-dashed green line is a model with redshift-dependent $R_e-R_h$ relation, and the dashed magenta line is a model with an evolving SMHM relation (see text). \emph{Bottom row:} The size functions of MSFGs, and S0 satellite galaxies assuming that, due to environmental effects, the MSFGs are transformed in S0 galaxies after a timescale $\Delta T_{transf}$, as labelled (MQGs shown in the inset).} 
    %\caption{\emph{Top row}: Variants to the benchmark "Frozen \& Stationary" (F+S) model shown in Figure \ref{fig:sizefunction_frozen}, and shown as a solid blue line here for reference. The dotted brown lines show a model with $\mu=1$. In the right panel, the cyan dot-dashed lines show the result of a model where galaxies are stripped according to the \citet{Smith+16_stripping} prescription, and their sizes are evolved along the $R_e-M_{\rm star}$ relation \citep{Hearin+19}. Models where the normalization of the $R_e-R_h$ relation increases with increasing redshift (green), or where the SMHM evolves (magenta) generate marginally improved size functions compared to the benchmark. \emph{Bottom row:} The size functions of MSFGs, MQGs and S0 satellite galaxies (inset), assuming that, due to environmental effects, the MSFGs are transformed in S0 galaxies after a timescale $\Delta T_{transf}$, as labelled.}  %The dotted red, dot-dashed orange and double-dashed gray lines show models with $\Delta T_{transf}=4$ Gyr, $\Delta T_{transf}$= 3 Gyr and $\Delta T_{transf}$ = 2 Gyr respectively.  Similarly, the number density of satellite MQGs is enriched by the injection of S0s in the quiescent population, as shown in the second panel. This model greatly improves on the predictions from F+S benchmark model for satellites MSFGs if $\Delta T_{transf}=3-4$ Gyr. }
    \label{fig:allmodels}
\end{figure*}

We will adopt as a reference throughout a basic ``Frozen \& Stationary" (F+S) model in which the SMHM relation does not depend on cosmic time and satellites do not evolve after infall. We will discuss below the possible impact of relaxing any of the assumptions in the F+S model.  
The left and right panels in Figure \ref{fig:sizefunction_frozen} show, respectively, the size function of starforming (MSFGs) and quiescent (MQGs) MGs extracted from SDSS and divided into central (orange diamonds) and satellite (blue triangles) galaxies. We compare these data with our F+S model (solid coloured lines). We first confirm the results of Paper I, our model provides an excellent match to the size function of central MGs. Here we show that, in addition, without any extra fine-tuning, the same model provides a good match also to the size function of satellite MGs, especially for the quenched population. It is also clear from Figure \ref{fig:sizefunction_frozen} that the vast majority of satellite MGs have been accreted at $z_{peak}<0.5$ (violet lines). In Figure \ref{fig:sizefunction_frozen_halobins} we further dissect the size function of the model MQGs in bins of host halo mass that are representative of low-mass groups (12.5$<\log M_h/M_{\odot}<13.3$), groups and low-mass clusters ($13.3<\log M_h/M_{\odot}<14$), and massive clusters ($\log M_h/M_{\odot}>14$). The mean size (arrows at the bottom) show a weak dependence on parent halo mass, with an increase in normalised size of $\Delta \gamma \lesssim 45\%$ (inset in Figure \ref{fig:sizefunction_frozen_halobins}), in line with what seen for central galaxies for which $\Delta \gamma \lesssim 55\%$ (insets in Figure~\ref{fig:sizefunction_frozen}). 
%{\color{green} Not sure about what follows. This is because more massive satellites, which live in more massive haloes at infall, also infall in more massive central halos. That's why I said originally that this is a cosmological effect that has to do with the assembly of the large scale structure$\longrightarrow$}. Although, as expected from the linearity in the $Re\propto R_h$ model, larger satellites tend to inhabit more massive haloes, the increase in mean size (red arrows at the bottom of each panel) is contained to a factor of $\Delta \bar{R}_e \lesssim 60\%$ or, in terms of mass-normalised size (Section 1), of $\Delta \gamma \lesssim 30\%$ {\color{green} This was considering only the three panels shown in Figure 2. In fact, when doing the gamma-Mh relation this becomes a bit higher, around 50-60\%...}. 
As emphasized by several groups \citep[e.g.,][]{Huertas-Company+13_environment,Sonnefeld+19}, the mass-normalized mean size $\gamma$ of central and satellite MGs has a weak dependence on host halo mass, amounting to $\Delta \gamma \lesssim 40\%$, when moving from field to clusters and after accounting for statistical measurements errors in host halo mass. Our models naturally generate a weak trend of mean size with halo mass mainly induced by the underlying assumption of a universal $R_e\propto R_h \propto M_h^{1/3}$ relation, in which the halo mass dependence is further washed out by dispersions in the relations and, in the case of satellites, by the stochastic assembly of haloes. As discussed by \citet{Shankar+14_sizes}, a weak dependence of the mean size with halo mass contrasts instead with some galaxy formation models, especially those characterised by strong disc instabilities. Figure~\ref{fig:sizefunction_frozen_halobins} also shows that most of the ``relic'' satellites (formed at $z_{peak}>1.5$) live today in massive clusters. Our results are largely independent of the specific inputs of the F+S model. For example, the top panels of Figure~\ref{fig:allmodels} show that similar size functions are generated when varying the quenching model (brown dotted lines, as labelled), or when allowing for some redshift evolution in the $R_e-R_h$ relation with $A_K\propto (1+z)^{0.2}$, still broadly allowed by the high-redshift data on the sizes of MGs (long-dashed, green lines, see Paper 2).

Despite the successes described so far, our model predictions are not perfect and present two main discrepancies from the data. Firstly, the predicted number density of satellite MSFGs tends to be progressively overestimated with respect to the SDSS data by a factor of $2-10$ below $R_e\sim10$ kpc (e.g., left top panel of Figure \ref{fig:allmodels}). Secondly, the model predicts a size function of MQGs very similar in shape to the measured one but shifted by $\sim 0.05$ dex towards lower sizes (see right panels of Figure \ref{fig:allmodels}). Despite being relatively small discrepancies, especially in the case of the MQG population, it is a non-trivial task to reconcile the models with the data by simply fine-tuning some of the input parameters. %{\color{green} delete? $\longrightarrow$--our model outputs are quite robust against variations in the input parameters, as mentioned above.} 
To prove this point, the magenta dashed line in the top right panel of Figure \ref{fig:allmodels} marks the outcome of a model in which we allow the input SMHM relation to vary with redshift. More specifically, we used a Markov Chain Monte Carlo algorithm \citep{Foreman-Mackey+13}
%\footnote{We used the package \texttt{emcee}, \citet{Foreman-Mackey+13}.}, 
which we run for 150,000 steps with 96 walkers with Gaussian priors centred on the mean of the posterior distributions shown in Appendix A of \citet{Grylls+20_satellites}. Although the resulting best-fit SMHM relation provides an improved fit to the low-size tail of the size function of MQGs, it still falls somewhat short at the high-size end and, more importantly, the implied stellar mass function, we verified, appears in stark disagreement with current data \citep{Davidzon+17,Kawinwanichakij+20}. Even a model in which we include stellar stripping at the rate suggested by \citet{Smith+16_stripping}, does not significantly alter the predicted size function of MQGs from the benchmark F+S model (cyan dashed line, top right panel of Figure \ref{fig:allmodels}). Simpler solutions to the (small) discrepancy in the predicted size function of MQG with respect to the data can be ascribed to, e.g., a possible overestimation of the sizes in satellite galaxies due to background subtraction effects, and/or a small deviation in the adopted $R_e-R_h$ relation in satellite galaxies with respect to their central counterparts. %{\color{red}*********}. 

As anticipated above, the most prominent discrepancy with the data lies in the overproduction of the number density of MSFGs, progressively increasing towards lower sizes (left panels of Figure \ref{fig:allmodels}).
Although part of the mismatch may in part also be caused by {incompleteness due to fiber collisions}\citep[e.g.,][]{Taylor+10}
%(e.g., {\color{green} move at the end of previous paragraph where asterisks are$\longrightarrow$ the number density of small galaxies may be underestimated in SDSS due to fiber collisions \citealt{Taylor+10}})
, in what follows we will only focus on the modelling side. Simply varying the relevant input parameters has no noticeable impact on the shape of the predicted size function of MSFGs (top left panel), thus calling for additional assumptions in the model. In the bottom panels of Figure \ref{fig:allmodels} we explore the impact of a physically-motivated hypothesis (e.g., \citealt{Cava+17_WINGS_S0s,Joshi+20}) in which MSFGs are morphologically transformed, without any change in size, 
%{\color{green} not sure the negligible effect on size is physically motivated...}, 
into massive lenticulars (MS0s) via the effect of the gas in the intra-group and intra-cluster media on a typical timescale of $\Delta T_{transf} \approx 2-4$ Gyr since $z_{peak}$ (coloured lines as labelled). This simple addition to our baseline model provides a nearly perfect match to the size function of MSFGs when adopting $\Delta T_{transf} \approx 3-4$ Gyr. The number of MS0 formed via this channel amounts, however, to only a few percent of the total population of MQGs (inset in bottom-left panel) and $\sim 10\%$ of the population of SDSS MS0 galaxies (bottom right panel), suggesting that the vast majority of MS0s may preferentially form before accreting in the cluster environment (e.g., \citealt{Hopkins09_diskmergers,Saha&Cortesi2019}), as opposed to what is seen at lower $M_{\rm star}$ observationally (e.g., \citealt{Desai+07}). %{\color{green} Allowing for star formation to occur for $\sim2$Gyrs after infall \citep{Wetzel+13}, following the
\section{Conclusions}
\label{sec:discussion}

In previous work we showed that assuming a universal $R_e-R_h$ relation provides an excellent match to the local size function of SDSS galaxies and to the strong size evolution of massive galaxies. 
In this Letter we further demonstrate that a basic ``frozen \& stationary'' model where (i) the SMHM and the $R_e-R_h$ relations remain unchanged since $z \sim 1.5$ and (ii) the environment does not affect galaxies after infall, predicts a local size function of massive satellite galaxies in good agreement with the data, particularly for massive quenched galaxies (MQGs). The same model generates an overall mild dependence of galaxy sizes on host halo mass for satellites , amounting to $\Delta \gamma\lesssim 45\%$, in agreement with observational studies \citep{Huertas-Company+13_environment}. Our results are robust against sensible (time) variations in the $R_e-R_h$ and/or SMHM relation, inclusion of stellar stripping, or variations in the quenching model. On the other hand, our model overpredicts the number density of massive star forming galaxies (MSFGs), especially at lower sizes. We find that by allowing for MSFGs to quench and transform into massive S0 galaxies in a timescale of $\Delta T_{transf}\approx 3-4$ Gyr, yields a nearly perfect match to the size function of MFGs. However, the fraction of S0 galaxies formed via the environmental channel would only amount to $\sim 10\%$ of the total number of massive S0s in SDSS, the vast majority of which must have preferentially formed in situ. The $R_e-R_h$ appears as a fundamental relation in regulating the size growth and environmental dependence of MGs.

\section*{Acknowledgements}
We thank the referee for a constructive report. FS acknowledges partial support from a Leverhulme Trust Research Fellowship. 
\section*{Data Availability}
Data will be shared upon request to the authors.
%The authors gratefully acknowledge the Gauss Centre for Supercomputing e.V. (www.gauss-centre.eu) and the Partnership for Advanced Supercomputing in Europe (PRACE, www.prace-ri.eu) for funding the MultiDark simulation project by providing computing time on the GCS Supercomputer SuperMUC at Leibniz Supercomputing Centre (LRZ, www.lrz.de). The CosmoSim database used in this paper is a service by the Leibniz-Institute for Astrophysics Potsdam (AIP).
%The MultiDark database was developed in cooperation with the Spanish MultiDark Consolider Project CSD2009-00064

%%%%%%%%%%%%%%%%%%%%%%%%%%%%%%%%%%%%%%%%%%%%%%%%%%
%\section*{Data Availability}
%The data used in this paper...

%%%%%%%%%%%%%%%%%%%% REFERENCES %%%%%%%%%%%%%%%%%%

% The best way to enter references is to use BibTeX:

\bibliographystyle{mnras}
\bibliography{papersize} % if your bibtex file is called example.bib

% Alternatively you could enter them by hand, like this:
% This method is tedious and prone to error if you have lots of references
%\begin{thebibliography}{99}
%\bibitem[\protect\citeauthoryear{Author}{2012}]{Author2012}
%Author A.~N., 2013, Journal of Improbable Astronomy, 1, 1
%\bibitem[\protect\citeauthoryear{Others}{2013}]{Others2013}
%Others S., 2012, Journal of Interesting Stuff, 17, 198
%\end{thebibliography}

%%%%%%%%%%%%%%%%%%%%%%%%%%%%%%%%%%%%%%%%%%%%%%%%%%

%%%%%%%%%%%%%%%%% APPENDICES %%%%%%%%%%%%%%%%%%%%%

%%%%%%%%%%%%%%%%%%%%%%%%%%%%%%%%%%%%%%%%%%%%%%%%%%

% Don't change these lines
\bsp	% typesetting comment
\label{lastpage}
\end{document}